\documentclass[%
 reprint,
 aps,
pra,
]{revtex4-2}
\usepackage{graphicx}
\usepackage{dcolumn}
\usepackage{bm}
\usepackage{color}
\usepackage{ulem}

\begin{document}


\title{Dynamical properties of Fermi-Fermi mixtures of dipolar and non-dipolar atoms}

\author{Takahiko Miyakawa}
\email{takamiya@auecc.aichi-edu.ac.jp}
\affiliation{
Department of Science Education, Aichi University of Education, Kariya 448-8542, Japan}
\author{Eiji Nakano}
\affiliation{
Department of Mathematics and Physics, Kochi University, Kochi 780-8520, Japan}
\author{Hiroyuki Yabu}
\affiliation{
Department of Physics, Ritsumeikan University, Kusatsu 525-8577, Japan}%

\date{\today}

\begin{abstract}
Dynamical properties of homogeneous Fermi-Fermi mixtures 
of dipolar and non-dipolar atoms are studied at zero temperature, where dipoles are polarized by an external field. 
We calculate the density-density correlation functions in a ring-diagram approximation 
and analyze the pole structure to obtain eigenfrequencies of collective excitations.  
We first determine stability phase diagrams for the mixtures available in experiments: 
$^{167}$Er~-~$^{173}$Yb, $^{167}$Er~-~$^{6}$Li, $^{161}$Dy~-~$^{173}$Yb, and $^{161}$Dy~-~$^{6}$Li systems, 
and show that the mixtures with larger mass imbalance tend to be more unstable. 
We then investigate the parameter dependence of an undamped zero sound with an anisotropic real dispersion relation  in the stable phase for the $^{161}$Dy~-~$^{173}$Yb mixture, 
and the speed of sound exhibits a critical angle of possible propagation with respect to the dipole polarization direction, 
above which the sound mode disappears in the particle-hole continuum. 
Since the sound mode is a coherent superposition of density fluctuations of dipolar and non-dipolar atoms, 
the existence of the sound mode, e.g., the value of the critical angle, is significantly affected by 
the inter-particle interaction through the density-density correlation between dipolar and non-dipolar atoms. 
We have also observed such an effect of the inter-particle interaction 
in the study of a linear response of density fluctuations to an external perturbation.  
\end{abstract}

\maketitle


\section{\label{sec:level1}Introduction}
One-component polarized dipolar Fermi gases have been realized experimentally 
using highly magnetic atoms of  
$^{161}$Dy~\cite{Lu2012}, $^{167}$Er~\cite{Aikawa2014a}, and $^{53}$Cr~\cite{Naylor2015}, 
respectively. 
In the study of such dipolar Fermi gases at low temperatures, 
the Fermi surface deformation is one of the most important quantum many-body phenomena, 
which was in fact observed in the experiment~\cite{Aikawa2014b,Chomaz2023}. 
The deformation was predicted theoretically prior to the experiment 
as a genuine quantum phenomenon 
originating from the exchange contribution of the anisotropic dipole-dipole interaction 
between identical dipolar fermions~\cite{Miyakawa2008,Sogo2009}, 
and further 
theoretical studies have revealed so far  
many interesting phenomena expected in such degenerate dipolar gases: 
an anisotropic zero sound propagation~\cite{Chan2010,Ronen2010}, 
anisotropic superfluids~\cite{Baranov2002,Zhao2010}, 
biaxial nematic phases~\cite{Fregoso2009}, 
topological superfluid phases~\cite{Cooper2009}, 
and density-wave phases~\cite{Yamaguchi2010,Sun2010}. 
Experimental researches for these phenomena, however, remain untouched 
because of the weakness of effective magnetic dipole-dipole interactions and 
the low number densities of these atomic gases achieved after cooling processes in trap.

In recent years a new progress has been made in experimental studies related to one-component polarized dipolar Fermi gases, that is, the realization of Fermi-Fermi gaseous mixtures of dipolar (magnetic) and non-dipolar (non-magnetic) atoms toward the investigation of mutual effects on their quantum many-body properties. 
To this end, it is essential to accomplish the quantum degeneracy 
and the inter-particle Feshbach resonance in the mixture, 
which have been realized by stages using various pairs of dipolar and non-dipolar Fermi atoms
including the experimental studies of $^{161}$Dy-$^{40}$K mixture 
by Inssbruck group~\cite{Ravensbergen2018,Ravensbergen2020,Ye2022}, 
$^{53}$Cr-$^{6}$Li by Firenze group~\cite{Ciamei2022a,Ciamei2022b}, 
and $^{167}$Er-$^{6}$Li by Kyoto group~\cite{Schafer2023a}. 
Incidentally, the quantum-degenerate Bose-Bose mixture of $^{168}$Er and $^{174}$Yb atoms 
has also been realized by Kyoto group~\cite{Schafer2023b}, 
suggesting that the quantum-degenerate mass-imbalanced Fermi-Fermi mixture of $^{167}$Er and $^{173}$Yb atoms is expected to be realized in near future. 

The advantageous points for experimental and theoretical studies of these mixtures 
lies in the selectivity of the mass ratio, 
and in the controllability of the number-density ratio 
and that of the inter-particle interaction strength between different species via Feshbach resonances.
In the preceding paper~\cite{Miyakawa2023}
on the study of collective excitations 
in homogeneous Fermi-Fermi mixtures of dipolar and non-dipolar atoms, 
we have found two different types of collective modes at zero temperature: 
one is the undamped zero sound characterized by an anisotropic dispersion relation 
and the other is an over-damped mode with purely imaginary frequencies; 
The zero sound may emerge to propagate within a restricted range of angle: $0\le \theta_{\bm q} \le \theta_{\bm q}^c$, 
where $\theta_{\bm q}$ denotes the angle between the dipole-polarization direction 
and the momentum of propagation $\hbar{\bm q}$, 
while the over-damped mode emerges complementarily in $\theta_{\bm q}^c \le \theta_{\bm q} \le \pi/2$. 
Here $\theta_{\bm q}^c$ is a critical angle at which the effective interaction in density-density channel vanishes; 
The effective interaction remains repulsive in $0\le \theta_{\bm q} \le \theta_{\bm q}^c$ 
to support the zero sound, 
while it becomes attractive in $\theta_{\bm q}^c \le \theta_{\bm q} \le \pi/2$. 
Furthermore, the over-damped mode turns into unstable one 
when the strength of inter-particle interaction exceeds some critical value. 
Since these collective modes are coherent superposition of dipolar and non-dipolar density fluctuations, 
these dynamical properties depend on the parameters such as 
the $s$-wave scattering length $a_s$ of  the inter-particle interaction, 
the mass ratio $r_m$, and the number-density ratio $r_n$, in a complex manner. 
In the present paper, we focus on the dynamical properties of the undamped zero sound entirely,  
and figure out in detail how its speed, critical angle, and amplitude depend on the parameters mentioned above, 
by taking samples of mixture of dipolar and non-dipolar atoms realized in experiments. 
We determine the stability phase diagram in the parameter space in advance, 
and then investigate dynamical properties of the zero sound in the stable phase. 
We also investigate the linear response of density fluctuations to an external perturbation. 
As a study of similar subject, the zero sound and instability in dilute nuclear matter are discussed in Fermi liquid theory~\cite{Kolomeitsev2016}.

This paper is organized as follows: 
In Sec.~II, 
we present the model of Fermi-Fermi mixtures of dipolar and non-dipolar atoms, 
and formulate the density-density correlation functions in a ring-diagram approximation. 
In Sec.~III, 
we draw the stability phase diagrams from the pole structure of the correlation function 
in cases of 
$^{167}$Er~-~$^{173}$Yb, $^{167}$Er~-~$^{6}$Li, $^{161}$Dy~-~$^{173}$Yb, $^{161}$Dy~-~$^{6}$Li mixtures. 
In Sec.~IV, using the correlation functions 
we calculate the dispersion relation and the speed of the undamped zero sound numerically 
to figure out their parameter dependence, 
and also investigate the induced density fluctuations of $^{161}$Dy~-~$^{173}$Yb mixtures
in the linear response to an impulsive perturbation. 
Here we stress that the role of the inter-particle interaction and the number-density ratio are of particular interest. 
Sec.~V is devoted to summary.

\section{Formalism}
To study the homogeneous gaseous mixtures of dipolar and non-dipolar Fermi atoms,
we employ the model Hamiltonian defined as follows: 
\begin{eqnarray}
\label{model}
 \hat{H} &=& \sum_{\bm k}\epsilon^0_{1\bm k} c_{1\bm k}^\dagger c_{1\bm k}+ \sum_{\bm k}\epsilon^0_{2\bm k} c_{2\bm k}^\dagger c_{2\bm k} \nonumber \\
  &+&\frac{1}{2}\sum_{\bm k , \bm k^\prime, \bm q}V_{dd}(\bm q)c_{1\bm k}^\dagger c_{1\bm k^\prime+\bm q}^\dagger c_{1\bm k^\prime} c_{1\bm k +\bm q} \nonumber \\
 &+&g\sum_{\bm k , \bm k^\prime, \bm q}c_{1\bm k}^\dagger c_{2\bm k^\prime+\bm q}^\dagger c_{2\bm k^\prime} c_{1\bm k +\bm q},
 \end{eqnarray}
where $\epsilon^0_{i\bm k}=\hbar^2 {\bm k}^2/2m_i$ ($i= 1$ or $2$) the kinetic energy of dipolar  or non-dipolar atoms. 
Accordingly, 
the annihilation and creation operators $c_{1\bm k}$ and $c^\dagger_{1\bm k}$ 
are for the dipolar Fermi atoms with the momentum $\hbar \bm k$, the mass $m_1$ and the dipole moment $\bm d$; 
Similarly, $c_{2\bm k}$ and $c^\dagger_{2\bm k}$ are 
for the non-dipolar Fermi atoms 
with the momentum $\hbar \bm k$ and the mass $m_2$. 
The dipoles are assumed to be polarized along the $z$-axis by an external field. 
The term including $V_{dd}$ is the dipolar interaction in the Fourier space, 
$V_{dd}({\bm q})=\frac{4\pi}{3}d^2(3\cos^2\theta_{\bm q}-1)$, 
where $\theta_{\bm q}$ is the angle between the momentum $\hbar {\bm q}$ and the dipole polarization direction.
The coupling constant $g$ is for the inter-particle interaction 
between dipolar and non-dipolar fermions, 
which is given by 
$g=2\pi\hbar^2 a_s/\mu$ 
with $a_s$ being the $s$-wave scattering length 
and $\mu = m_1m_2/(m_1+m_2)$ the reduced mass.
We take the volume of the system to be unity in this paper.

To implement the perturbative treatment on the basis of Hartree-Fock (HF) ground state,
it is necessary to introduce  
the particle and hole operators, 
$a_{i\bm k}$ and $b_{i\bm k}$, defined by 
\begin{eqnarray}
  c_{i\bm k} &=& \theta(\epsilon_{i\bm k} - \epsilon_{i F}) a_{i\bm k} +\theta(\epsilon_{i F}-\epsilon_{i\bm k})b_{i-\bm k}^\dagger, \\
  c_{i\bm k}^\dagger &=& \theta(\epsilon_{i\bm k} - \epsilon_{iF}) a_{i\bm k}^\dagger +\theta(\epsilon_{iF}
  -\epsilon_{i\bm k})b_{i-\bm k}, 
  \end{eqnarray}
where $\epsilon_{i\bm k}$ denotes the HF single-particle energy for dipolar ($i=1$) or non-dipolar ($i=2$) atoms, 
and $\epsilon_{i F}$($i=1,2$) the corresponding Fermi energies.  
In terms of these particle and hole operators, 
the Hamiltonian (\ref{model}) is rewritten as 
\begin{eqnarray*}
  \hat{H}&=&E_0+\sum_{i=1,2} \sum_{\bm k} \epsilon_{i\bm k} \theta(\epsilon_{i\bm k} - \epsilon_{i F}) a^\dagger_{i\bm k}a_{i\bm k} \\
  &-&\sum_{i=1,2}\sum_{\bm k} \epsilon_{i\bm k}\theta(\epsilon_{iF}-\epsilon_{i\bm k}) b^\dagger_{i\bm k}b_{i\bm k} \nonumber \\
 &+&\frac{1}{2}\sum_{\bm k,\bm k^\prime, \bm q} V_{dd}(\bm q)N\left( c_{1\bm k}^\dagger c_{1\bm k^\prime+\bm q}^\dagger c_{1\bm k^\prime} c_{1\bm k +\bm q} \right) \nonumber \\
 &+& g\sum_{\bm k,\bm k^\prime, \bm q} N\left( c_{1\bm k}^\dagger c_{1\bm k+\bm q} c_{2\bm k^\prime+\bm q}^\dagger c_{2\bm p^\prime}  \right),
\end{eqnarray*}
where $E_0$ is the HF ground state energy and 
the symbol $N$ denotes the normal ordering for particle and hole operators.

\begin{figure}[h]
\includegraphics[width=8.5cm]{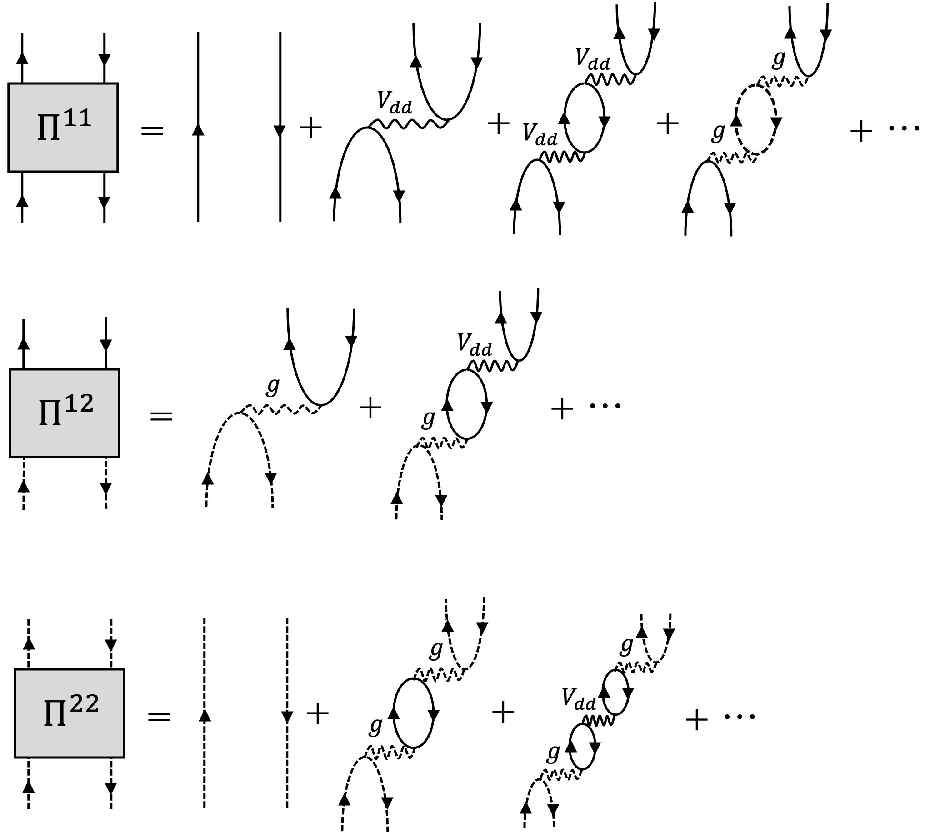}
\caption{\label{ringdiagram} Ring-diagrams of dipolar and non-dipolar Fermi atoms. The solid and dashed lines represent dipolar and non-dipolar fermions, respectively. The solid and dashed wavy-lines represent the dipolar and inter-particle interactions, respectively. }
\end{figure}

To find the collective excitations in density-fluctuation channels, 
we will evaluate general density-density correlation functions for the mixture defined as follows: 
\begin{eqnarray}
   i\hbar \Pi^{ij}(\bm q,t)= \qquad\qquad\qquad\qquad\qquad\qquad\qquad\quad \nonumber \\
  \sum_{\bm k, \bm k^\prime} \left\langle T\left[ c_{Hi\bm k}^\dagger (t)c_{Hi\bm k +\bm q}(t)
  c_{ Hj\bm k^\prime +\bm q}^\dagger (0)c_{Hj\bm k^\prime} (0)\right] \right\rangle
\end{eqnarray}
where $i, j= 1,2$ and the symbol $T$ denotes the time-ordering product. 
The operators $c_{Hi\bm k}(t)$ and $c_{Hi\bm k}^\dagger (t)$ are Heisenberg operators, 
and the expectation value $\langle \cdots \rangle$ is taken with respect to the exact Heisenberg ground state. 
In order to analyze the above correlation functions, 
we evaluate $\Pi^{ij}$ perturbatively 
on the basis of the HF ground state using a ring-diagram approximation~\cite{NP,FW,Miyakawa2023}, 
where the ring-diagrams of dipolar and non-dipolar fermions are mixed 
via the inter-particle interaction as depicted in Figure~\ref{ringdiagram}. 

In this approximation, 
the correlation functions are calculated for the wavevector ${\bm q}$ and frequency $\omega$ to be 
\begin{eqnarray}
 \Pi^{11}(\bm q, \omega)&=&\frac{\Pi^{11}_0}{1-V_{dd}\Pi^{11}_0-g^2\Pi^{11}_0\Pi^{22}_0},\\
 \Pi^{12}(\bm q, \omega)&=&\Pi^{21}(\bm q, \omega)=\frac{g\Pi^{11}_0\Pi^{22}_0}{1-V_{dd}\Pi^{11}_0-g^2\Pi^{11}_0\Pi^{22}_0},\\
 \Pi^{22}(\bm q, \omega)&=&\frac{(1-V_{dd}\Pi^{11}_0)\Pi^{22}_0}{1-V_{dd}\Pi^{11}_0-g^2\Pi^{11}_0\Pi^{22}_0},
\end{eqnarray}
where 
$\Pi^{ii}_0(\bm q,\omega)$ is the single-loop polarization functions with respect to the HF ground state, 
\begin{eqnarray}
\label{polarizationfunc}
  \Pi^{ii}_0(\bm q,\omega)&=&
  \sum_{\bm k} 
  \left[\frac{(1-f_{i\bm k+\bm q})f_{i\bm k}}{\hbar\omega + \epsilon_{i\bm k}-\epsilon_{i\bm k +\bm q}+i\eta}
  \right.
  \nonumber 
  \\   
 && \qquad\qquad -\left. \frac{f_{i\bm k+\bm q}(1-f_{i\bm k})}{\hbar\omega + \epsilon_{i\bm k}-\epsilon_{i\bm k +\bm q}-i\eta}
  \right]. 
\end{eqnarray}
Here $f_{i\bm k}$ ($i=1,2$) is the Fermi-Dirac distribution function in the HF approximation. 

The dispersion relation of the collective excitations is obtained 
from the poles of the retarded correlation functions $\Pi^{ij}_R(\omega)$; 
the eigenfrequency $\omega_{\bm q}$ of the collective excitations are determined 
from the eigenvalue equation:
\begin{equation}
\label{dispersion}
  1=\left[V_{dd}(\bm q)+g^2\Pi^{22}_{0R}(\bm q, \omega_{\bm q})\right]\Pi^{11}_{0R}(\bm q, \omega_{\bm q})
\end{equation}
where 
$\Pi^{ii}_{0R}(\bm q, \omega_{\bm q})={\rm Re}\Pi^{ii}_{0}(\bm q, \omega_{\bm q})
+i {\rm sgn}\, \omega_{\bm q} \,{\rm Im}\Pi^{ii}_{0}(\bm q, \omega_{\bm q})$ \cite{FW}. 
It should be noted that in the eigenvalue equation (\ref{dispersion}) the factor 
$\left[V_{dd}(\bm q)+g^2\Pi^{22}_{0R}(\bm q, \omega_{\bm q})\right]$ 
plays a role of the effective density-density interaction, 
and in general when it is positive the undamped zero sound mode may emerge.   

The polarization function $\Pi^{11}_{0R}$ can be evaluated using the variational ansatz 
for the distribution function of dipolar Fermi gases \cite{Miyakawa2008}:
\begin{equation}
   f_{1\bm k}=\theta\left(k_{1F}^2-\frac{1}{\beta}(k_x^2+k_y^2)-\beta^2k_z^2\right) ,
\end{equation}
where $k_{1F}=(6\pi^2 n_1)^{1/3}$, 
and the $n_1$ is the number-density of dipolar Fermi gas;
the parameter $\beta$ ($\beta <1$) characterizes Fermi surface  (non-spherical) deformation 
caused by the exchange energy of dipolar interaction. 

In this variational approximation, 
the HF single-particle energy $\epsilon_{1\bm k}$ of dipolar Fermi gas is given by
\begin{equation}
   \epsilon_{1\bm k} = \epsilon(0)+\frac{\hbar^2}{2m_1}\lambda^2\left(\frac{1}{\beta}(k_x^2+k_y^2)+\beta^2k_z^2 \right),
\end{equation}
where $\epsilon(0)$ and $\lambda^2$ represent the energy shift and 
the curvature of the single-particle energy (an effective mass), respectively. 
The parameters $\beta$, $\epsilon(0)$, and $\lambda$ 
included in the above equation can be evaluated 
in the variational approximation method~\cite{Nishimura2021}, 
and they are shown to depend on the dimensionless dipolar interaction strength $k_{1F}a_{dd}$ 
with $a_{dd}=m_1d^2/3\hbar^2$ the dipolar length. 
Then, the real and imaginary parts of $\Pi^{11}_{0R}(\bm q, \omega)$ become
\begin{widetext}
\begin{eqnarray}
\label{repart11}
  {\rm Re}\,\Pi^{11}_{0R}&=&C_{11}\left[
  -1+\frac{k_{1F}}{2q_\beta}\left\{1-\left(\frac{\omega}{\lambda^2v_{1F}q_\beta}-\frac{q_\beta}{2k_{1F}}\right)^2 \right\}\ln\left|\frac{1+\left(\frac{\omega}{\lambda^2v_{1F}q_\beta}-\frac{q_\beta}{2k_{1F}}\right)}
  {1-\left(\frac{\omega}{\lambda^2v_{1F}q_\beta}-\frac{q_\beta}{2k_{1F}}\right)} \right|  \right.\nonumber\\
  &-&\left. \frac{k_{1F}}{2q_\beta}\left\{1-\left(\frac{\omega}{\lambda^2v_{1F}q_\beta}+\frac{q_\beta}{2k_{1F}}\right)^2 \right\}\ln\left|\frac{1+\left(\frac{\omega}{\lambda^2v_{1F}q_\beta}+\frac{q_\beta}{2k_{1F}}\right)}
  {1-\left(\frac{\omega}{\lambda^2v_{1F}q_\beta}+\frac{q_\beta}{2k_{1F}}\right)} \right| 
  \right],
\end{eqnarray}
and
\begin{eqnarray}
\label{impart11}
   {\rm Im}\,\Pi^{11}_{0R}=
   \left\{\begin{array}{l}
   -\frac{C_{11} k_{1F}}{2q_\beta} \left[1-\left(\frac{\omega}{\lambda^2v_{1F}q_\beta}-\frac{q_\beta}{2k_{1F}}\right)^2 
   \right] ;
   \quad 1\leq \frac{q_\beta}{2k_{1F}}\; {\rm and}\; \frac{q_\beta}{2k_{1F}}-1 \leq \frac{\omega}{\lambda^2 v_{1F} q_\beta} \leq \frac{q_\beta}{2k_{1F}}+1\\
    -\frac{C_{11} k_{1F}}{2q_\beta} \left[1-\left(\frac{\omega}{\lambda^2v_{1F}q_\beta}-\frac{q_\beta}{2k_{1F}}\right)^2 
   \right] ;
   \quad 0 \leq\frac{q_\beta}{2k_{1F}} <1\; {\rm and}\; 
   1-\frac{q_\beta}{2k_{1F}} \leq \frac{\omega}{\lambda^2 v_{1F} q_\beta} \leq \frac{q_\beta}{2k_{1F}}+1\\
   -\frac{C_{11} \omega}{\lambda^2 v_{1F}q_\beta} ;
   \quad 0 \leq\frac{q_\beta}{2k_{1F}} <1\; {\rm and}\; 
   0 \leq \frac{\omega}{\lambda^2 v_{1F} q_\beta} < 1-\frac{q_\beta}{2k_{1F}} \\
   0; \quad \;{\rm otherwise}, 
\end{array} 
   \right.
\end{eqnarray}
\end{widetext}
respectively, 
where we have defined 
$C_{11}=\frac{m_1k_{1F}}{4\pi^2\hbar^2\lambda^2}$, 
$q_\beta=(q^2_x/\beta+q^2_y/\beta+\beta^2q_z^2)^{1/2}$, 
and $v_{1F}=\hbar k_{1F}/m_1$.
 
In the same way, we obtain the real and imaginary parts of $\Pi^{22}_{0R}(\bm q, \omega)$:
\begin{widetext}
\begin{eqnarray}
\label{repart22}
  {\rm Re}\,\Pi^{22}_{0R}&=&C_{22}\left[
  -1+\frac{k_{2F}}{2q}\left\{1-\left(\frac{\omega}{v_{2F}q}-\frac{q}{2k_{2F}}\right)^2 \right\}\ln\left|\frac{1+\left(\frac{\omega}{v_{2F}q}-\frac{q}{2k_{2F}}\right)}
  {1-\left(\frac{\omega}{v_{2F}q}-\frac{q}{2k_{2F}}\right)} \right|  \right.\nonumber\\
  &-&\left. \frac{k_{2F}}{2q}\left\{1-\left(\frac{\omega}{v_{2F}q}+\frac{q}{2k_{2F}}\right)^2 \right\}\ln\left|\frac{1+\left(\frac{\omega}{v_{2F}q}+\frac{q}{2k_{2F}}\right)}
  {1-\left(\frac{\omega}{v_{2F}q}+\frac{q}{2k_{2F}}\right)} \right| 
  \right],
\end{eqnarray}
and,
\begin{eqnarray}
\label{impart22}
   {\rm Im}\,\Pi^{22}_{0R}=
   \left\{\begin{array}{l}
   -\frac{C_{22} k_{2F}}{2q} \left[1-\left(\frac{\omega}{v_{2F}q}-\frac{q}{2k_{2F}}\right)^2 
   \right] ;
   \quad 1\leq \frac{q}{2k_{2F}}\; {\rm and}\; \frac{q}{2k_{2F}}-1 \leq \frac{\omega}{v_{2F} q} \leq \frac{q}{2k_{2F}}+1\\
    -\frac{C_{22} k_{2F}}{2q} \left[1-\left(\frac{\omega}{v_{2F}q}-\frac{q}{2k_{2F}}\right)^2 
   \right] ;
   \quad 0 \leq\frac{q}{2k_{2F}} <1\; {\rm and}\; 
   1-\frac{q}{2k_{2F}} \leq \frac{\omega}{v_{2F} q} \leq \frac{q}{2k_{2F}}+1\\
   -\frac{C_{22}\omega}{v_{2F}q} ;
   \quad 0 \leq\frac{q}{2k_{2F}} <1\; {\rm and}\; 
   0 \leq \frac{\omega}{v_{2F} q} < 1-\frac{q}{2k_{2F}} \\
   0; \quad \;{\rm otherwise}, 
\end{array} 
   \right.
\end{eqnarray}
\end{widetext}
respectively, 
where $q=|\bm q|$, $C_{22}=\frac{m_2k_{2F}}{4\pi^2\hbar^2}$, and 
$v_{2F}=\hbar k_{2F}/m_2$  
where $k_{2F}=(6\pi^2 n_2)^{1/3}$ 
with $n_2$ being the number density of non-dipolar atoms.

\section{Stability of Fermi-Fermi mixtures}
In this section 
we determine the stability phase diagram 
for experimentally realizable dipolar and non-dipolar Fermi mixtures: 
$\rm ^{167}Er$-$\rm ^{173}Yb$, 
$\rm ^{167}Er$-$\rm ^{6}Li$,  
$\rm ^{161}Dy$-$\rm ^{173}Yb$, and 
$\rm ^{161}Dy$-$\rm ^{6}Li$.
From the eigenvalue equation~(\ref{dispersion}), 
the stability condition for a given $\bm q$ is obtained as  
\begin{equation}
\label{stability}
  1\geq \left[V_{dd}(\bm q)+g^2\Pi^{22}_{0R}(\bm q, 0)\right] \Pi^{11}_{0R}(\bm q, 0). 
\end{equation}
It should be noted that the anisotropic dipole-dipole interaction $V_{dd}(\bm q)$ 
takes the minimum value at $\theta_{\bm q}=\frac{\pi}{2}$, i.e., $V_{dd}(\bm q) =-\frac{4\pi}{3}d^2$, 
which is negative and independent of the magnitude of momentum.  
When $\omega=0$, $\Pi^{11}_{0R}$ and $\Pi^{22}_{0R}$ are negative and those absolute values decrease monotonically with increasing $q$ at $\theta_{\bm q}=\frac{\pi}{2}$. it is likely that the system becomes most unstable against 
the homogeneous density fluctuations ($q\rightarrow 0$) 
in the direction perpendicular to the dipole polarization one, i.e., $\theta_{\bm q}=\frac{\pi}{2}$. 
Therefore, the stability condition of the mixture becomes
\begin{equation}
\label{stabilitycondition}
   1\geq \frac{2}{\pi}\frac{k_{1F}a_{dd}}{\lambda^2}+\frac{1}{\pi^2}\left(\sqrt{r_m}+\frac{1}{\sqrt{r_m}}\right)^2
   \frac{r_n^{1/3}(k_{1F}a_s)^2}{\lambda^2},
\end{equation}
where $r_m=m_2/m_1$ and $r_n=n_2/n_1$ are the mass ratio and the number-density ratio, respectively.
It should be noted that a dipolar and non-dipolar mixture becomes unstable if the stability condition is not satisfied, irrespective of the sign of the scattering length $a_s$ as shown in Eq.~(\ref{stabilitycondition}). In the unstable region for positive $a_s$, the mixed gas undergoes a phase separation between dipolar and non-dipolar gases, whereas for negative $a_s$ it collapses to a dense phase~\cite{Houbiers1997}. In what follows, we consider the stability condition only for positive $a_s$ (repulsive inter-particle interaction).

\begin{figure}[h]
\includegraphics[width=8.5cm]{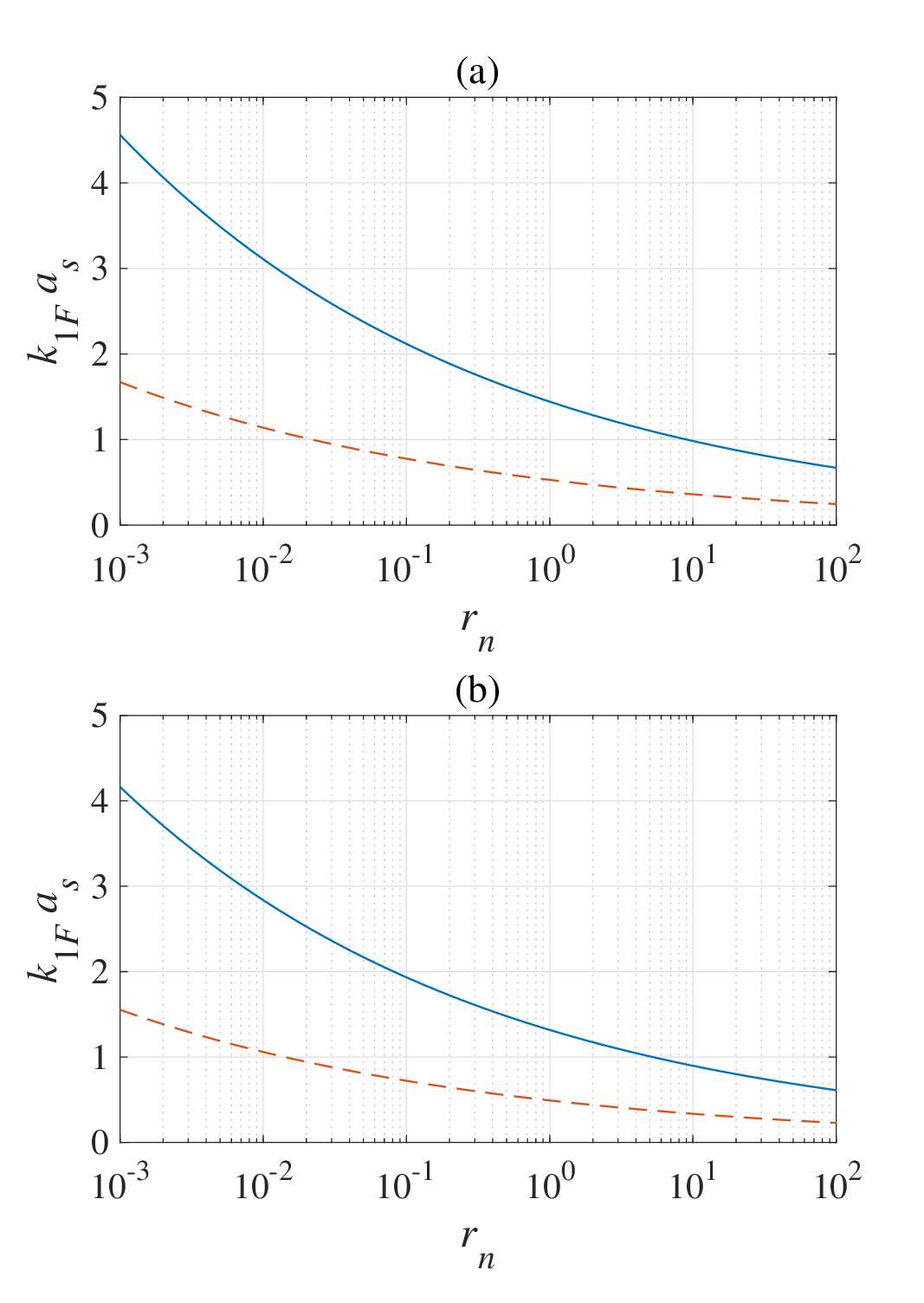}
\caption{\label{stabilityfig} Stability diagrams obtained from (\ref{stabilitycondition}) 
in the plane of the dimensionless inter-particle scattering length $k_{1F}a_s$ and 
the number-density ratio $r_n=n_2/n_1$
for mixtures of 
(a) $\rm ^{167}Er$-$\rm ^{173}Yb$ (solid line) 
and $\rm ^{167}Er$-$\rm ^{6}Li$ (dashed line) for $k_{1F}a_{dd}=0.250$, 
and (b) $\rm ^{161}Dy$-$\rm ^{173}Yb$ (solid line) and $\rm ^{161}Dy$-$\rm ^{6}Li$ (dashed line) 
for $k_{1F}a_{dd}=0.482$. Each system is stable in the region below line.}
\end{figure}

Figure \ref{stabilityfig} shows stability diagrams in the $r_n$-$k_{1F}a_s$ plane 
for dipolar and non-dipolar mixtures: 
(a) the $\rm ^{167}Er$-$\rm ^{173}Yb$ and  $\rm ^{167}Er$-$\rm ^{6}Li$ for $k_{1F}a_{dd}=0.250$, 
and (b) the $\rm ^{161}Dy$-$\rm ^{173}Yb$ and $\rm ^{161}Dy$-$\rm ^{6}Li$ 
for $k_{1F}a_{dd}=0.482$~\cite{Footnote1}. 
The $\rm ^{167}Er$-$\rm ^{173}Yb$ and $\rm ^{167}Er$-$\rm ^{6}Li$ mixtures are stable in the region below the solid and dashed lines in Fig.~\ref{stabilityfig}~(a), respectively, 
while $\rm ^{161}Dy$-$\rm ^{173}Yb$ and $\rm ^{161}Dy$-$\rm ^{173}Yb$ are stable 
below the solid and dashed lines in Fig.~\ref{stabilityfig}~(b). 
From these figures, 
larger mass-imbalanced mixtures are found to be more unstable. 
It should be understood by the existence of the pre-factor $(\sqrt{r_m}+\frac{1}{\sqrt{r_m}})^2$ 
in the stability condition~(\ref{stabilitycondition}). 
Since the magnetic moment of the Dy atom is about 10/7 times as large as that of the Er atom, 
the dipolar interaction of Dy atoms is stronger than Er atoms;
it explains why the critical line of the Dy atom is pushed downward 
in comparison with that of the Er atom as shown in Figs.~\ref{stabilityfig} (a) and (b).

Here we make a comment on the stability condition of the phase separation for positive $a_s$; according to theoretical studies of two-component repulsive atomic Fermi gases~\cite{TCRFG}, for instance, the quantum fluctuations may cause a phase separation with lower values of $k_{1F}a_{s}$ than the mean-field result obtained in this study. Nevertheless, in the following sections, we will develop a ring-diagram approximation for collective excitations upon the mean-field ground state in a consistent manner.

\section{Dynamical properties of the $\rm ^{161}Dy$-$\rm ^{173}Yb$ mixture}
In this section, 
we focus on the system of $\rm ^{161}Dy$-$\rm ^{173}Yb$ mixture 
which provides $k_{1F}a_{dd}=0.482$ and $r_m=1.00746$, 
and investigate its dynamical properties as varying parameters of $k_{1F}a_s$ and $r_n$ 
within ranges available in experiments. 
The qualitative behavior of dynamical properties are expected 
not to change significantly for other mixtures of mass-imbalanced dipolar and non-dipolar atoms~\cite{Footnote2}.

\subsection{Undamped zero sound}
The eigenvalue equation (\ref{dispersion}) generally admits a complex eigenfrequency solution: 
$\omega_{\bm q} = \Omega_{\bm q}-i\Gamma_{\bm q}$,  
which corresponds to the collective excitations 
of the dipolar and non-dipolar Fermi mixtures. 
As is shown in our previous paper~\cite{Miyakawa2023}, 
the undamped zero-sound mode with $\Gamma_{\bm q}=0$ appears only when the mixture is stable.

\begin{figure}[t]
\includegraphics[width=8.5cm]{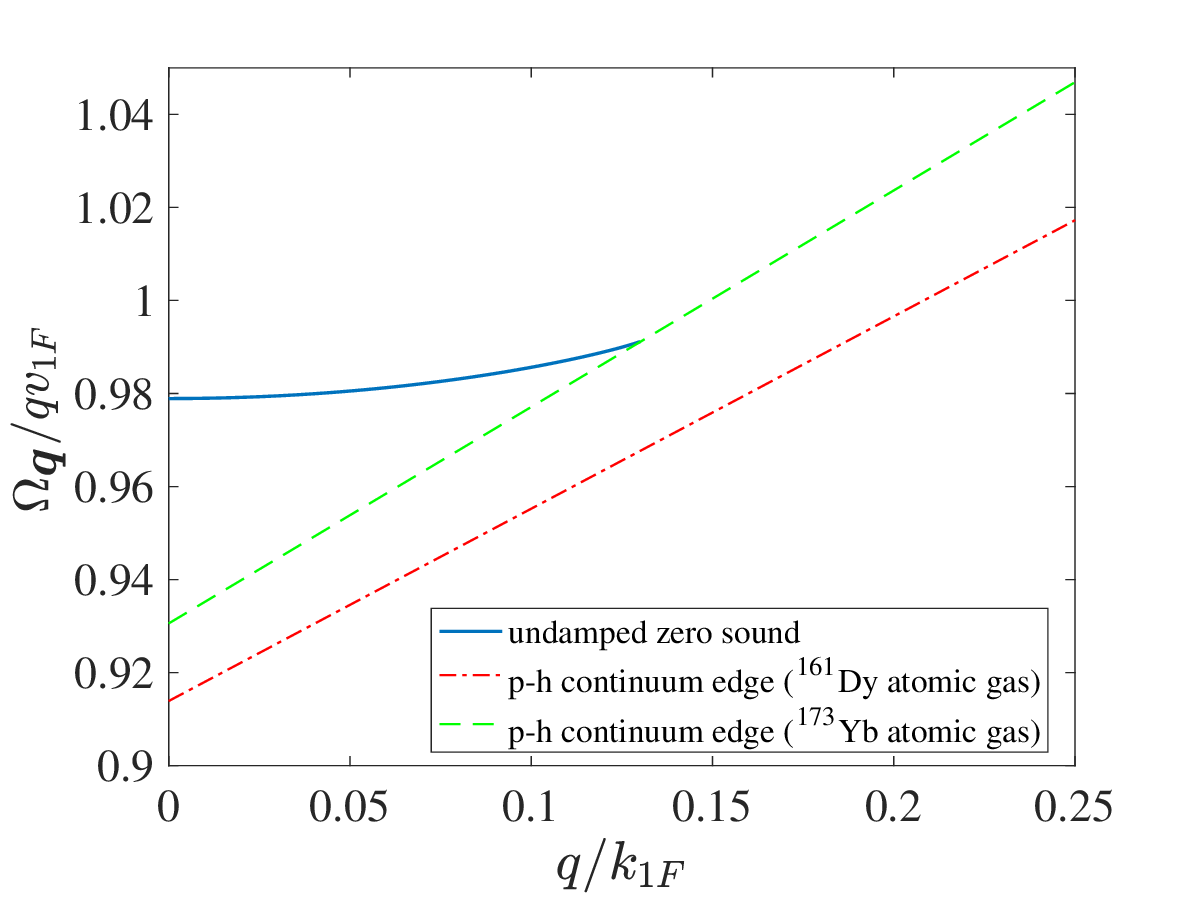}
\caption{\label{momentum} Eigenfrequency of undamped zero sound mode 
(scaled with the momentum $\Omega_{\bm q}/qv_{1F}$) 
for the momentum transfer $\bm q = (0, 0, q)$ 
of the $^{161}$Dy - $^{173}$Yb 
($k_{1F}a_{dd}=0.482$ and $r_m=1.00746$) for $k_{1F}a_s=1.30$, $r_n=1.00$ (solid line). 
Dot-dashed and dashed lines represent the edges of the incoherent particle-hole continuum 
of $\rm ^{161}Dy$ and $\rm ^{173}Yb$ atomic gases, respectively.}
\end{figure}

Figure~\ref{momentum} shows the eigenfrequency $\Omega_{\bm q}$ of the undamped zero sound mode  
as a function of the transfer $\bm q = (0, 0, q)$, i.e., along the dipole polarization direction. 
In the figure the eigenfrequency seems to encounter the edge of incoherent particle-hole continuum solution 
at a point of some finite $q$, 
but a careful analysis of the solution shows that the amplitude of the sound mode vanishes exactly at this point.

\begin{figure}[h]
\includegraphics[width=8.5cm]{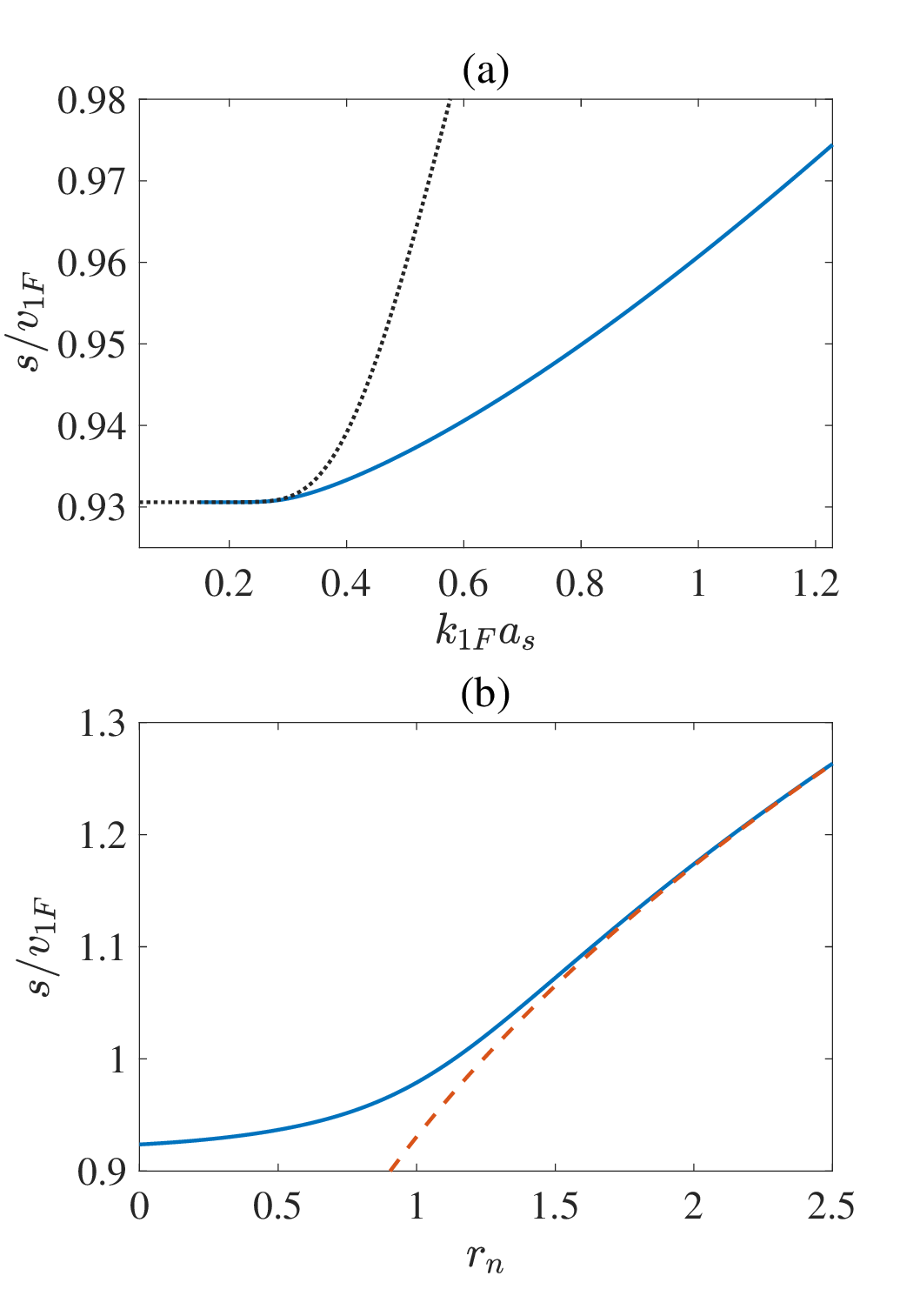}
\caption{\label{soundspeed} Speeds of the undamped zero sound  (sollid line) 
with $\theta_{\bm q}=0$ of the $^{161}$Dy - $^{173}$Yb mixture ($k_{1F}a_{dd}=0.482$ and $r_m=1.00746$) 
as a function of (a) $k_{1F}a_s$ for $r_n=1.00$, and (b) $r_n$ for $k_{1F}a_s=1.30$. 
The dotted line in (a) represents the speed of sound in the weak-coupling regime given by Eq.~(\ref{weakcoupling1}). 
The dashed line in (b) represents the edge of incoherent particle-hole continuum of  $^{173}$Yb atomic gas.}
\end{figure}

In Figure~\ref{soundspeed}, 
we show the speed of the undamped  zero sound 
defined by 
\begin{equation}
\label{s1}
\displaystyle s=\lim_{q \to 0}\frac{\Omega_{\bm q}}{q} 
\end{equation}
in the case of $\theta_{\bm q}=0$, i.e., for the sound propagating along the dipole polarization direction. 
In calculations of the speed of sound in what follows, 
we take the long-wavelength limit ($q\to 0$) of the real parts of $\Pi_{0R}^{11}$ and $\Pi_{0R}^{22}$ 
with the ratio $\omega/q$ fixed, which result in 
%
\begin{eqnarray*}
	{\rm Re}\,\Pi^{11}_{0R} &\rightarrow& C_{11}\left[ \frac{\omega}{\lambda^2v_{1F}q \alpha_\beta(\theta_{\bm q})}
	\ln\left|\frac{1+\frac{\omega}{\lambda^2v_{1F}q \alpha_\beta(\theta_{\bm q})}}
	{1-\frac{\omega}{\lambda^2v_{1F}q \alpha_\beta(\theta_{\bm q})}}\right| -2\right], 
	\\
	{\rm Re}\,\Pi^{22}_{0R} &\rightarrow& C_{22}\left[ \frac{\omega}{v_{2F}q}
	\ln\left|\frac{1+\frac{\omega}{v_{2F}q}}{1-\frac{\omega}{v_{2F}q}}\right| -2\right], 
\end{eqnarray*}
where 
\[
\alpha_{\beta}(\theta_{\bm q}) = (\sin^2{\theta_{\bm q}}/\beta+\beta^2 \cos^2{\theta_{\bm q}})^{1/2}.
\]

As shown in Fig.~\ref{soundspeed} (a), 
the speed of sound (solid line) increases monotonically with the value of $k_{1F}a_s$. 
The dotted line in Fig.~\ref{soundspeed}(a) represents the speed of sound 
in the weak-coupling regime, which is given in an analytical form as 
\begin{widetext}
\begin{equation}
\label{weakcoupling1}
  s = s_{20}\left[1+2 \exp\left\{-2-\frac{2\pi^2\lambda^2}{\left(\sqrt{r_m}+\frac{1}{\sqrt{r_m}}\right)^2r_n^{1/3}(k_{1F}a_s)^2}
  \left(\frac{2}{\Phi^{11}(s_{20})}-\frac{4k_{1F}a_{dd}}{\pi\lambda^2}\right)\right\}
  \right],
\end{equation}
\end{widetext}
where $s_{20}=r_n^{1/3}v_{1F}/r_m$  is the speed of sound at particle-hole continuum edge of non-dipolar atoms, and
\[
	\Phi^{11}(s)= \frac{s}{\lambda^2\beta v_{1F}}
	\ln\left|\frac{1+\frac{s}{\lambda^2\beta v_{1F}}}{1-\frac{s}{\lambda^2\beta v_{1F}}}\right| -2.
\]
%
%
%
As shown in Fig.~\ref{soundspeed} (a), 
it agrees with the numerical results (solid line) 
in the region of small values of $k_{1F}a_s$.
Moreover, 
as shown in Fig.~\ref{soundspeed} (b), 
the speed of sound (solid line) increases monotonically with $r_n$, 
and it approaches the edge of incoherent particle-hole continuum of $\rm ^{173}Yb$ atomic gas (dashed line) 
as $r_n$ increases.

\begin{figure}[b]
\includegraphics[width=8.5cm]{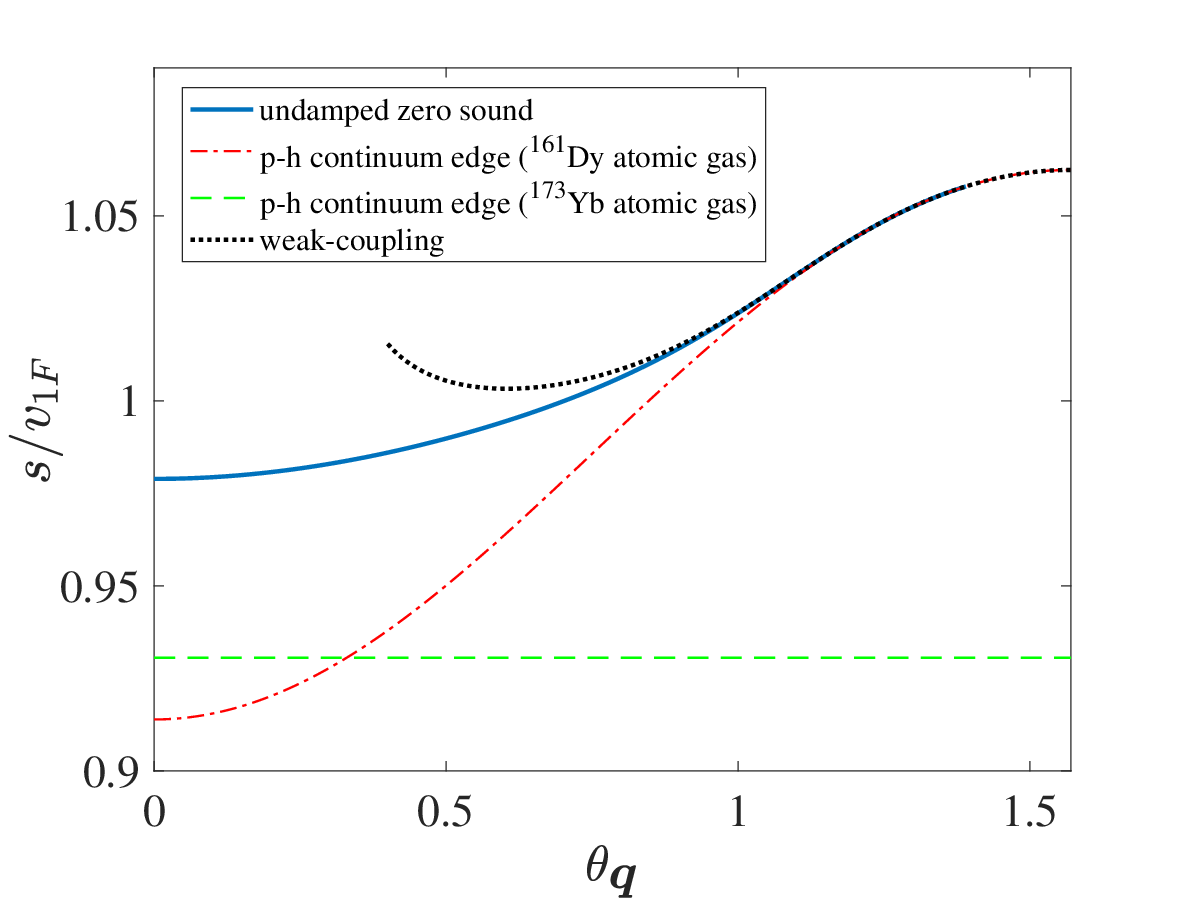}
\caption{\label{angle} Anisotropic speed of undamped zero sound of the $^{161}$Dy - $^{173}$Yb mixture 
($k_{1F}a_{dd}=0.482$ and $r_m=1.00746$), 
as a function of the angle $\theta_{\bm q}$ 
for $k_{1F}a_s=1.30$ and  $r_n=1.00$ (solid line). 
Dot-dashed and dashed lines represent the edges of incoherent particle-hole continuum 
of $^{161}$Dy  and $^{173}$Yb atomic gases, respectively. 
The dotted line is the speed of sound in the weak-coupling regime Eq.~(\ref{weakcoupling2}).}
\end{figure}

Figure~\ref{angle} shows the anisotropy of the speed of zero sound, 
i.e., its dependence on the angle $\theta_{\bm q}$. 
Dot-dashed and dashed lines represent the edges of incoherent particle-hole continuum 
of $^{161}$Dy and $^{173}$Yb atomic gases, respectively.
As the angle $\theta_{\bm q}$ gets close to $\pi/2$, 
the speed of zero sound approaches 
the edge of incoherent particle-hole continuum of $^{161}$Dy atomic gas. 
In this case, the speed of zero sound can be approximated by that in the weak-coupling regime:
\begin{widetext}
\begin{equation}
\label{weakcoupling2}
  s=s_{10}\left[1+2\exp\left\{-2-\frac{2}{ \frac{2k_{1F}a_{dd}}{\pi\lambda^2}(3\cos\theta^2_{\bm q}-1)
    +\left(\sqrt{r_m}+\frac{1}{\sqrt{r_m}}\right)^2\frac{(k_{1F}a_s)^2r_n^{1/3}}{2\pi^2\lambda^2}\Phi^{22}(s_{10})}\right\}\right]
\end{equation}
\end{widetext}
where $s_{10}=\lambda^2 \alpha_\beta(\theta_{\bm q})v_{1F}$ is the speed of sound at particle-hole continuum edge of dipolar atoms, and
\[
	\Phi^{22}(s)= \frac{r_m s}{r_n^{1/3}v_{1F}}
	\ln\left|\frac{1+\frac{r_m s}{r_n^{1/3}v_{1F}}}{1-\frac{r_m s}{r_n^{1/3}v_{1F}}}\right| -2.
\]
As shown in Fig.~\ref{angle}, the numerical results (solid line) agree well with 
those from the weak-coupling approximation (\ref{weakcoupling2}) (dotted line)
in the region where $1.0 \lesssim \theta_{\bm q}$.

In the absence of the contact interaction ($a_s=0$), 
or in the case of a pure dipolar Fermi gas, 
the undamped zero sound exists only in the region, 
$0 \leq \theta_{\bm q} \leq \arccos(1/\sqrt{3})\simeq 0.9553$, 
where the dipolar interaction $V_{dd}(\bm q)$ becomes repulsive. 
In the case of mixtures, on the other hand, 
the undamped zero sound exists only 
when the effective density-density interaction in Eq.~(\ref{dispersion}) becomes repulsive, i.e.,  
$V_{dd}({\bm q})+g^2 {\rm Re}\Pi_{0R}^{22}({\bm q},\Omega_{\bm q})>0$.
Since the eigenfrequency $\Omega_{\bm q}$ encounters the particle-hole continuum edge of the dipolar atoms 
as the critical angle $\theta_{\bm q}^c$ is approached from below, the eigenfrequency near the critical angle in the long-wavelength limit $q\to 0$ becomes $\Omega_{\bm q}=\lambda^2 \alpha_\beta(\theta_{\bm q})v_{1F} q$.
Thus, the critical angle is determined by  the condition 
\[
\lim_{q\to 0} \left[V_{dd}(\theta^c_{{\bm q}})+g^2 {\rm Re}\Pi_{0R}^{22}({\bm q},\lambda^2 \alpha_\beta(\theta_{\bm q}^c) v_{1F}q)\right]=0
\]

\begin{figure}[h]
\includegraphics[width=8.5cm]{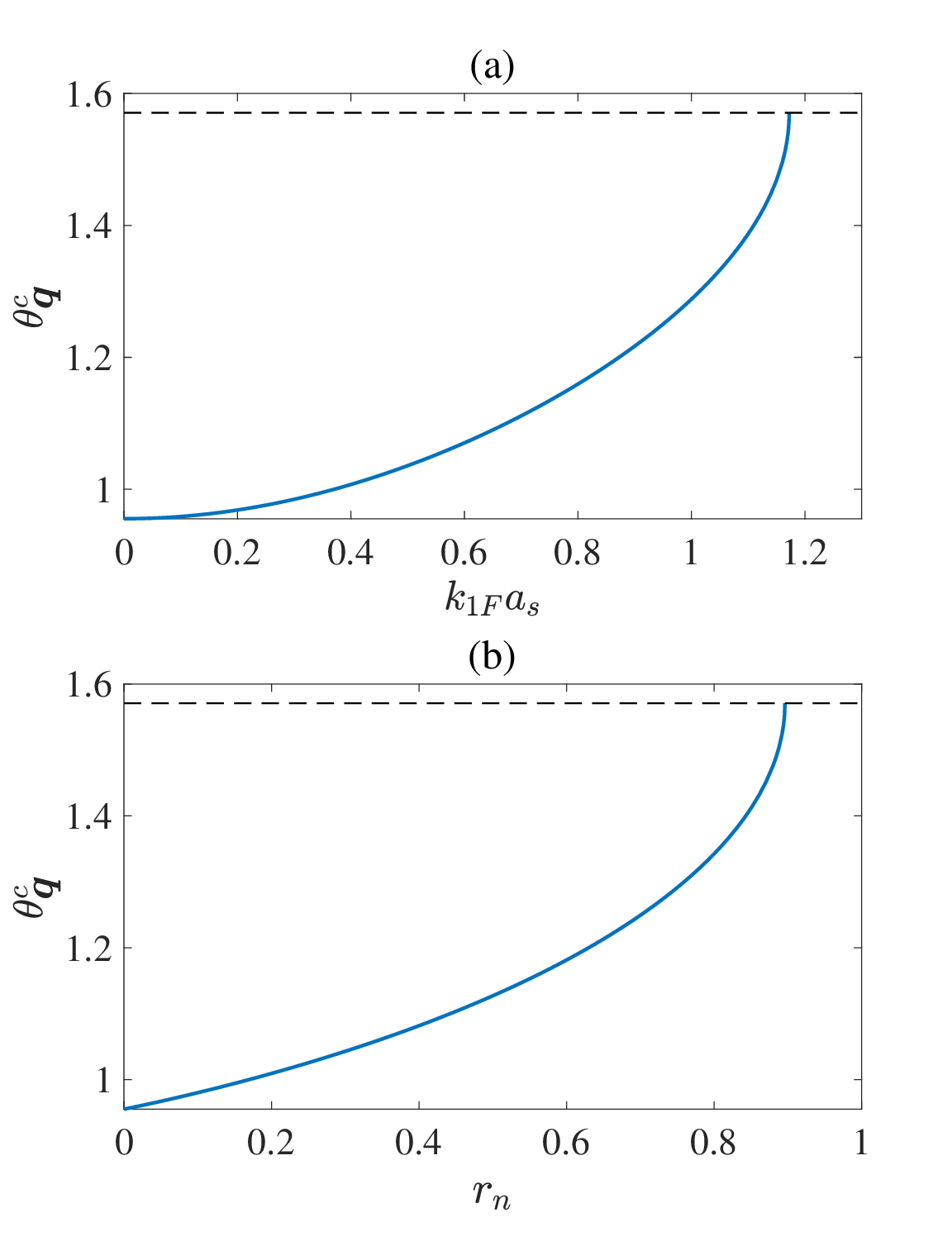}
\caption{\label{criticalangle} Critical angles (solid line) of the $^{161}$Dy - $^{173}$Yb mixture 
($k_{1F}a_{dd}=0.482$ and $r_m=1.00746$) as a function of (a) $k_{1F}a_s$ for $r_n=1.00$ and (b) $r_n$ for $k_{1F}a_s=1.30$. The dashed straight lines correspond to $\theta_{\bm q}^c=\pi/2$.}
\end{figure}

Figure~\ref{criticalangle} shows the critical angle 
of the $^{161}$Dy - $^{173}$Yb mixture as functions of 
(a) $k_{1F}a_s$ for $r_n=1.00$ and 
of (b) $r_n$ for $k_{1F}a_s=1.30$, respectively. 
As shown in Fig.~\ref{criticalangle}, 
the critical angles start from its pure dipolar Fermi gas limit, i.e., 
$\theta_{\bm q}^c=\arccos(1/\sqrt{3})$, 
and increase monotonically both with $k_{1F}a_s$ and $r_n$, 
due to the effect of the inter-particle interaction between dipolar and non-dipolar atoms, 
until $k_{1F}a_s$ exceeds $1.17$ for $r_n=1.00$ 
in Fig.~\ref{criticalangle}(a) and 
$r_n$ exceeds $0.90$ for $k_{1F}a_s=1.30$ in Fig.~\ref{criticalangle}(b), 
respectively.

The measurements of collective excitation spectra in atomic gases have been performed using fixed momentum two-photon Bragg spectroscopy~\cite{BraggSpectroscopy}, which is designed to make an arbitrary frequency and momentum transfer to density fluctuations. Using the experimental technique, the anisotropic properties of the undamped zero sound of dipolar and non-dipolar Fermi-Fermi mixtures discussed in this subsection can be observed.

\subsection{Density fluctuations induced by an impulsive perturbation}
Let us now turn to a discussion of linear response of $^{161}$Dy - $^{173}$Yb mixtures 
to an impulsive perturbation that is expressed by
\begin{equation}
\label{perturbation}
   \hat{H}_{ex}(t) = \int d^3r \left\{\hat{n}_1(\bm r, t)+\hat{n}_2(\bm r, t)\right\}U^{ex}(\bm r, t)
\end{equation}
where $U^{ex}(\bm r, t)=U_0^{ex}e^{i\bm q \cdot \bm r}\delta(t)$. 
The corresponding induced density fluctuations are given by 
\begin{eqnarray*}
   \delta n_i (\bm r, t)&=&\sum_{j=1}^{2}\delta n_{ij}(\bm r, t)  \\
  \mbox{with} \ \ 
  \delta n_{ij}(\bm r, t) &=& {\rm Re}\left[ U_0^{ex}
  e^{i\bm q \cdot \bm r}\int \frac{d\omega}{2\pi} e^{-i\omega t}\Pi_R^{ij}(\bm q, \omega) \right],
\end{eqnarray*}
where $\delta n_{ij}$ ($i,j=1,2$) represents the density fluctuation of the $i$-th atomic gas 
induced by the perturbation affecting the $j$-th atomic gas. 
Using the results obtained in previous sections, 
the density fluctuations $\delta n_{ij}$ induced by the undamped zero sound are calculated to be 
%
%
\begin{eqnarray}
   \delta n_{11}&=&U_0^{ex} \frac{\Pi_{0R}^{11}}
   {2F^\prime(\bm q, \Omega_{\bm q})}\sin(\bm q \cdot \bm r-\Omega_{\bm q}t), \\
   \delta n_{12}&=&\delta n_{21}=U_0^{ex} \frac{g\Pi_{0R}^{11}
   \Pi_{0R}^{22}}{2F^\prime(\bm q, \Omega_{\bm q})}\sin(\bm q \cdot \bm r-\Omega_{\bm q}t), \\
   \quad \nonumber\\
    \delta n_{22}&=&U_0^{ex} \frac{(1-V_{dd}\Pi_{0R}^{11})\Pi_{0R}^{22}}
   {2F^\prime(\bm q, \Omega_{\bm q})}\sin(\bm q \cdot \bm r-\Omega_{\bm q}t),
\end{eqnarray}
where
\[
   F^\prime(\bm q, \Omega_{\bm q}) 
   = -V_{dd} \left.\frac{\partial \Pi_{0R}^{11}}{\partial \omega}\right|_{\Omega_{\bm q}}
   -g^2 \left.\frac{\partial (\Pi_{0R}^{11}\Pi_{0R}^{22})}{\partial \omega}\right|_{\Omega_{\bm q}}.
\]
These derivatives of the polarization functions are given by
\begin{widetext}
	\begin{equation}
		\frac{\partial \Pi_{0R}^{11}}{\partial \omega}=\frac{C_{11}}{\lambda^2 v_{1F}q_\beta}\frac{k_{1F}}{q_\beta}
		\left[\left(\frac{\omega}{\lambda^2 v_{1F}q_\beta}+\frac{q_\beta}{2k_{1F}}\right)
		\ln\left|\frac{1+\left(\frac{\omega}{\lambda^2 v_{1F}q_\beta}+\frac{q_\beta}{2k_{1F}}\right)}
		{1-\left(\frac{\omega}{\lambda^2 v_{1F}q_\beta}+\frac{q_\beta}{2k_{1F}}\right)}\right|
		-\left(\frac{\omega}{\lambda^2 v_{1F}q_\beta}-\frac{q_\beta}{2k_{1F}}\right)
		\ln\left|\frac{1+\left(\frac{\omega}{\lambda^2 v_{1F}q_\beta}-\frac{q_\beta}{2k_{1F}}\right)}
		{1-\left(\frac{\omega}{\lambda^2 v_{1F}q_\beta}-\frac{q_\beta}{2k_{1F}}\right)}\right|
		\right],
	\end{equation}
	\begin{equation}
		\frac{\partial \Pi_{0R}^{22}}{\partial \omega}=\frac{C_{22}}{v_{2F}q}\frac{k_{2F}}{q}
		\left[\left(\frac{\omega}{v_{2F}q}+\frac{q}{2k_{2F}}\right)
		\ln\left|\frac{1+\left(\frac{\omega}{v_{2F}q}+\frac{q}{2k_{2F}}\right)}
		{1-\left(\frac{\omega}{v_{2F}q}+\frac{q}{2k_{2F}}\right)}\right|
		-\left(\frac{\omega}{v_{2F}q}-\frac{q}{2k_{2F}}\right)
		\ln\left|\frac{1+\left(\frac{\omega}{v_{2F}q}-\frac{q}{2k_{2F}}\right)}
		{1-\left(\frac{\omega}{v_{2F}q}-\frac{q}{2k_{2F}}\right)}\right|
		\right].
	\end{equation}
\end{widetext}
Getting all together, we obtain the total density fluctuation of the $i$-th atomic gas as 
\begin{equation}
   \delta n_{i} = A_{i} \sin(\bm q \cdot \bm r-\Omega_{\bm q}t)
\end{equation}
where the amplitudes $A_{1}$ and $A_2$ are given by
\begin{eqnarray}
\label{Amp1}
   A_{1} &=& U_0^{ex} \frac{\Pi_{0R}^{11}+g\Pi_{0R}^{11}\Pi_{0R}^{22}}{2F^\prime(\bm q, \Omega_{\bm q})},\\
\label{Amp2}
   A_{2} &=& U_0^{ex} \frac{g\Pi_{0R}^{11}\Pi_{0R}^{22}+(1-V_{dd}\Pi_{0R}^{11})\Pi_{0R}^{22}}{2F^\prime(\bm q, \Omega_{\bm q})}, 
\end{eqnarray}
respectively. 
In the long-wavelength limit, 
the derivatives of $\Pi_{0R}^{11}$ and $\Pi_{0R}^{22}$ become 
%
\begin{eqnarray*}
	\frac{\partial\,\Pi^{11}_{0R}}{\partial \omega} &=& \frac{C_{11}}{\lambda^2 v_{1F}q_\beta}
	\left[\ln\left|\frac{1+\frac{\omega}{\lambda^2v_{1F}q_\beta}}{1-\frac{\omega}{\lambda^2v_{1F}q_\beta}}\right|
	+\frac{\frac{2\omega}{\lambda^2v_{1F}q_\beta}}{1-\left(\frac{\omega}{\lambda^2v_{1F}q_\beta}\right)^2} \right], \\
	\frac{\partial\,\Pi^{22}_{0R}}{\partial \omega} &=& \frac{C_{22}}{v_{2F}q}
	\left[\ln\left|\frac{1+\frac{\omega}{v_{2F}q}}{1-\frac{\omega}{v_{2F}q}}\right|
	+\frac{\frac{2\omega}{v_{2F}q}}{1-\left(\frac{\omega}{v_{2F}q}\right)^2} \right],
\end{eqnarray*}
%
respectively. 
From the above equations 
both of the amplitudes are shown to be zero when $q=0$.

\begin{figure}[h]
\includegraphics[width=8.5cm]{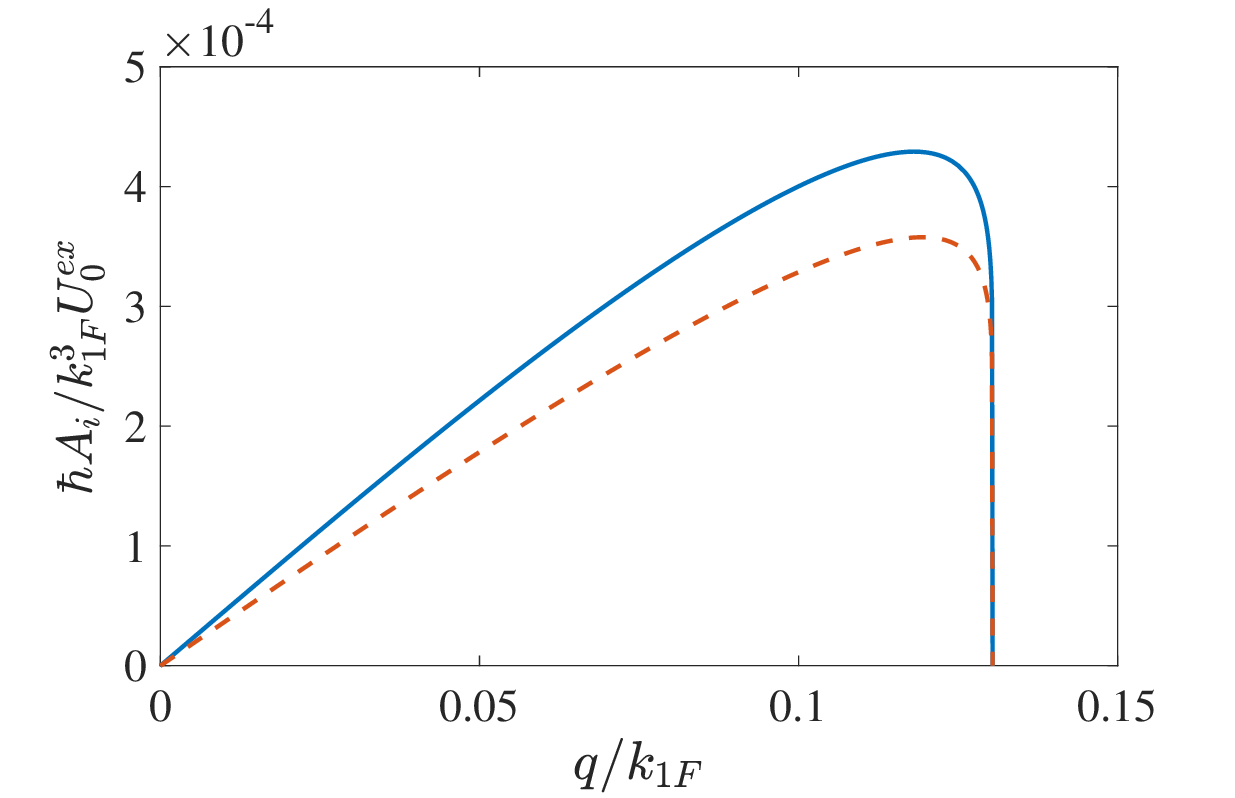}
\caption{\label{amplitudes} Amplitudes of density fluctuations of the $^{161}$Dy - $^{173}$Yb mixture  
where $k_{1F}a_{dd}=0.482$ and $r_m=1.00746$ induced by the impulsive perturbation 
$\hat{H}_{ex}(t) $ with the momentum transfer $\bm q =(0, 0, q)$ for $k_{1F}a_s=1.30$ and $r_n=1.00$. 
The solid and dashed lines show $\hbar A_1/ k_{1F}^3 U_0^{ex}$ and $\hbar A_2/ k_{1F}^3 U_0^{ex}$, respectively.}
\end{figure}

Figure~\ref{amplitudes} shows 
the amplitudes of density fluctuations $A_{1,2}$ in the $^{161}$Dy - $^{173}$Yb mixture 
induced by the impulsive perturbation $\hat{H}_{ex}(t)$ 
with the transfer $\bm q =(0, 0, q)$ for $k_{1F}a_s=1.30$ and $r_n=1.00$. 
The amplitudes gradually increases with $q$ up to a certain value, 
and beyond it they fall abruptly to be zero again at the point 
where the eigenfrequency of the undamped zero sound touches 
the incoherent particle-hole continuum edge of $^{173}{\rm Yb}$ atomic gas 
as shown in Fig.~\ref{momentum}.

\begin{figure}[h]
\includegraphics[width=8.5cm]{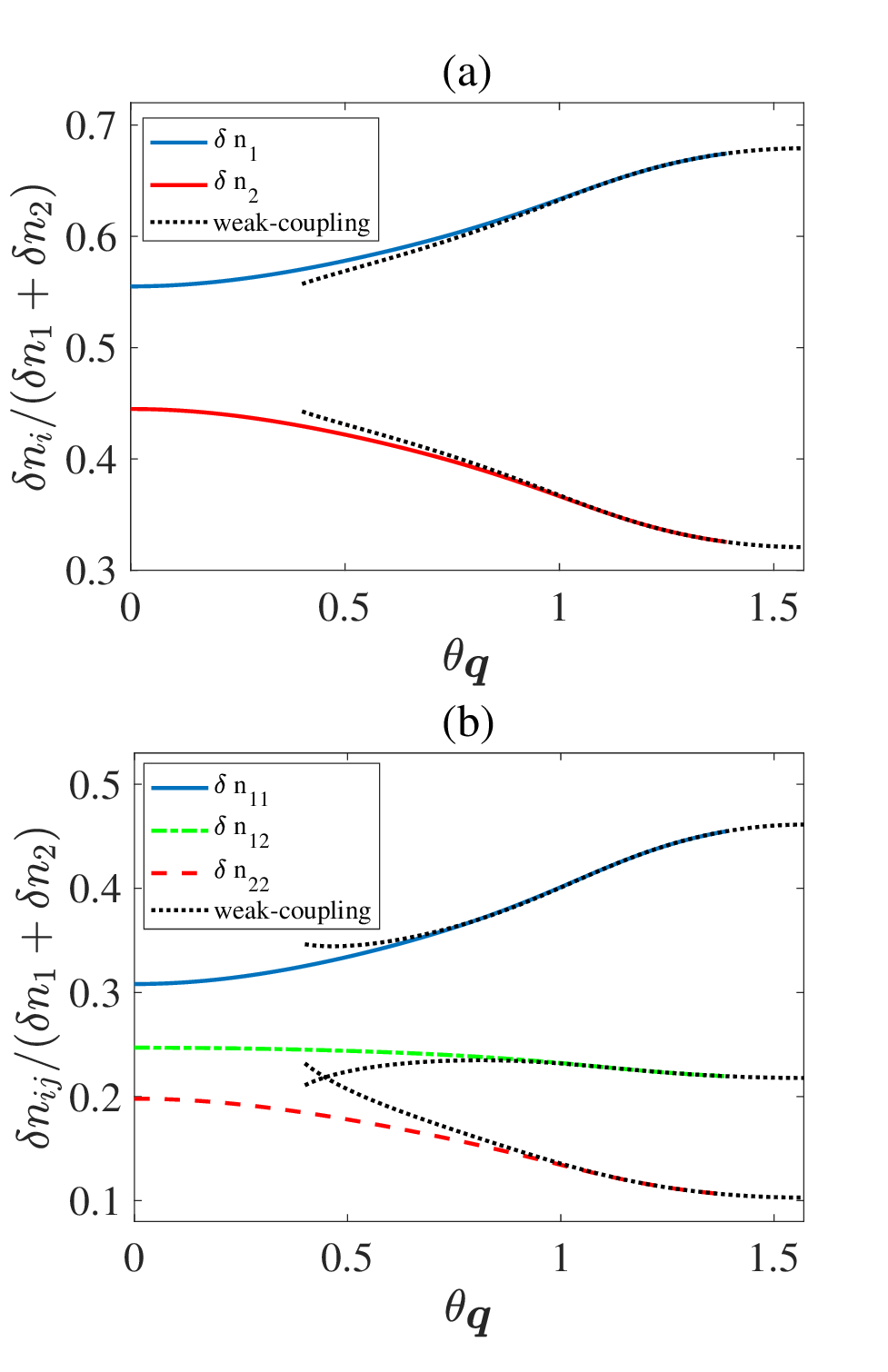}
\caption{\label{fluctuations} Relative density fluctuations of the $^{161}$Dy - $^{173}$Yb mixture 
where $k_{1F}a_{dd}=0.482$ and $r_m=1.00746$ in the limit of $q\to 0$ 
as a function of $\theta_{\bm q}$ for $k_{1F}a_s=1.30$ and $r_n=1.00$. 
Figures (a) and (b) show $\delta n_i/(\delta n_1+\delta n_2)$ ($i=1,2$) 
and $\delta n_{ij}/(\delta n_1 + \delta n_2)$ ($i,j=1,2$), respectively. 
The dotted lines in (a) and (b) represent results in the weak-coupling regime 
with the speed of sound given by Eq.~(\ref{weakcoupling2}).}
\end{figure}

Figure~\ref{fluctuations} shows the relative density fluctuations of the $^{161}$Dy - $^{173}$Yb mixture:
 (a) $\delta n_i/(\delta n_1+\delta n_2)$ and (b) $\delta n_{ij}/(\delta n_1+\delta n_2)$, 
in the limit of $q \to 0$ as a function of $\theta_{\bm q}$ for $k_{1F}a_s=1.30$ and $r_n=1.00$. 
As shown in Fig.~\ref{fluctuations}~(a), 
the density fluctuations of $^{161}{\rm Dy}$ atoms tend to be more dominant 
than those of $^{173}{\rm Yb}$ atoms  with increasing $\theta_{\bm q}$. 
The result is understood from the behavior in Fig.~\ref{angle} that the eigenfrequency of the undamped zero sound approaches the incoherent particle-hole continuum edge of $^{161}{\rm Dy}$ with increasing $\theta_{\bm q}$; that is the sound mode consists mostly of particle-hole pair states of $^{161}{\rm Dy}$ atoms just below the eigenfrequency of the sound mode.
Fig.~\ref{fluctuations}~(b) shows  
the relative fraction of $\delta n_{ij}$ in the total induced density fluctuation.
Similar to the results in Fig.~\ref{angle}, 
the numerical results in Figs.~\ref{fluctuations} (a) and (b) also agree well with those 
in the weak-coupling regime with the speed of sound given by Eq.~(\ref{weakcoupling2}) (dotted line) around the momentum angle region 
$1.0 \lesssim \theta_{\bm q}$. 

The impulsive perturbation can be realized using the technique of the short Bragg pulse in Ref.~\cite{BraggPulse}. The {\it in~situ} measurement of subsequent oscillations after a Bragg pulse can reveal the frequencies and amplitudes of collective oscillations of the dipolar and non-dipolar Fermi-Fermi mixtures.

\newpage

\section{Summary}
In this paper, 
we have studied the stability and dynamical properties 
of homogeneous dipolar and non-dipolar Fermi-Fermi mixtures at zero temperature. 
We have obtained the density-density correlation functions of the mixtures 
in the ring-diagram approximation, 
and analyzed the eigenvalue equations of the collective excitations. 
We have obtained the stability diagrams 
of the $^{167}$Er~-~$^{173}$Yb, $^{167}$Er~-~$^{6}$Li, $^{161}$Dy~-~$^{173}$Yb and $^{161}$Dy~-~$^{6}$Li mixtures, 
and found that the mixtures of larger mass imbalance tend to be more unstable. 
We have also investigated the eigenfrequency, 
the speed of undamped zero sound, and the density fluctuations 
of the $^{161}$Dy~-~$^{173}$Yb mixture
in the impulsive perturbation method; 
the results are summarized in Figs.~\ref{momentum}-\ref{fluctuations}. 
These results show that the inter-particle interaction between $^{161}$Dy and $^{173}$Yb atoms 
has a significant effect on the angle dependence of the sound propagation with respect to dipolar polarization direction, 
and also on the linear response of external perturbation
through the density-density correlation between $^{161}$Dy and $^{173}$Yb atoms.
These results can be experimentally observed using the Bragg spectroscopy technique \cite{BraggSpectroscopy, BraggPulse}.

In this paper, we obtained the excitation spectrum of undamped zero sound and density fluctuations by the impulsive perturbation in the $^{161}$Dy and $^{173}$Yb mixture, assuming repulsive inter-particle interaction ($a_s > 0$). It should be noted that the excitation spectrum depends only on the magnitude of scattering length $a_s$ as shown in Eq.~(\ref{dispersion}), whereas the amplitude of density fluctuations depends on the sign of $a_s$ as shown in Eqs.(\ref{Amp1}) and (\ref{Amp2}); we present the results only for positive $a_s$ in the subsection~IV-B. In the case of negative $a_s$, which we do not investigate in this study, one can expect the Cooper pairing between dipolar and non-dipolar atoms if the system is in the stable region of density fluctuations.
\\

\begin{acknowledgments}
This work was supported by Grants-in-Aid for Scientific Research through Grant No. 21K03422, provided by JSPS.
\end{acknowledgments}





\begin{thebibliography}{99}
\bibitem{Footnote1} The values of $k_{1F}a_{dd}$ discussed here are relatively larger than typical ones which are the order of a few percent. These values are reproduced by using the value of the central density of a dipolar Fermi gas with $~ 5\times 10^6$ particles in a hormonic oscillator potential with a frequency $500$Hz.
\bibitem{Footnote2} The ring-diagram approximation contains only one particle-hole pair excitations; it is valid for weak couplings: $k_{1F}a_{dd} \ll 1$ and $k_{1F}a_{s} \ll 1$. When the couplings are not weak, it is necessary to take into account multi-pair excitations too. However, dynamical properties of collective excitation in the long wavelength and low-frequency limit can be explained by the ring-diagram approximation, at least qualitatively, since single pair excitations are of importance in this case~\cite{NP}.
\end{thebibliography}
\end{document}